\documentclass{aastex701}
\usepackage{graphicx}%
\usepackage{amsmath,amssymb,amsfonts}%
\usepackage{amsthm}
\usepackage{mathrsfs}%
\usepackage{xcolor}%
\usepackage{textcomp}%
\usepackage{manyfoot}%
\usepackage{array}
\usepackage{xcolor}
\usepackage{tablefootnote}
\usepackage{tablefootnote}
\usepackage{booktabs}%
\usepackage{algorithm}%
\usepackage{algorithmicx}%
\usepackage{algpseudocode}%
\usepackage{listings}%
\usepackage{amsmath}
\usepackage{adjustbox}
\usepackage{tabularx}
\usepackage{cleveref}
\usepackage{tikz}
\usepackage{lipsum}
\graphicspath{{/plot/}}
\usepackage{etoolbox}
\usepackage{multirow}
\usepackage{rotating}
\usepackage{comment}
\usepackage{gensymb}
\usepackage{textcomp}
\usepackage{float}

\usepackage{amssymb}
\usepackage{appendix}

\usepackage{array}
\usepackage{caption}

\makeatletter
\providecommand{\@reinserts}{}
\makeatother
\shortauthors{Saha et al.}
\begin{document}

\title{Precise and Rapid Parameter Inference of Kilonova with Conditional Variational Autoencoder}
\correspondingauthor{Surojit Saha}
\email{surojitsaha@gapp.nthu.edu.tw}

\author[orcid=0000-0000-0000-0001]{Surojit Saha}
\affiliation{Institute of Astronomy, National Tsing Hua University, Hsinchu City, Taiwan, R.O.C}
\email{surojitsaha@gapp.nthu.edu.tw}

\author[orcid=0000-0000-0000-0002]{Albert K.H Kong} 
\affiliation{Institute of Astronomy, National Tsing Hua University, Hsinchu City, Taiwan, R.O.C}
\affiliation{Institute of Space Engineering, National Tsing Hua University, Hsinchu City, Taiwan, R.O.C}
\email{akong@gapp.nthu.edu.tw}


\begin{abstract}
The coalescence of binary neutron stars in the GW170817 event led to the generation of gravitational waves, accompanied by the electromagnetic counterpart known as a kilonova (KN). Since then, it has been a prime topic of interest, as it has provided much insight into multi-messenger astronomy. Apart from existing methods for parameter estimation, we propose an alternative technique for it, utilizing the strength and flexibility of a conditional variational autoencoder. Publicly available light curves are used as training data, conditioning on the corresponding physical parameters for a chosen model; after training, we carry out rapid parameter inferences. As this approach approximates the likelihood through variational inference, it yields results more efficiently. Through this innovative approach, we demonstrated that the total time, from training to parameter inference, is under $\approx3$h. We showed that for a given KN light curve, we can rapidly perform parameter inference based on the required model.

\end{abstract}


\section{Introduction}
\label{sec:intro}

Parameter inference constitutes an integral part of various astrophysical scenarios where comprehensive knowledge of the underlying physical parameters related to an astrophysical phenomenon is immensely important, consequently aiding in drawing conclusions between the observation and the proposed model. In the context of parameter estimation, detailed maximum likelihood estimation (MLE)\citep{1979ApJ...228..939C,1979ApJ...230..274W,refId0}, Markov Chain Monte Carlo (MCMC) \citep{c7ccf55c-6141-3386-abc6-c7d63d8427cb,annurev:/content/journals/10.1146/annurev-astro-082214-122339,article}, and Bayesian inference \citep{Loredo1990,Loredo2013,chattopadhyay2014statistical,2015bmps.book.....A,Thrane_Talbot_2019,2023arXiv230204703E} calculations have always been the preferred approach; however, this often requires longer completion times and consumes additional computational resources. One of the effective ways to reduce both is to leverage machine learning (ML) techniques. At a glance, it might appear that ML techniques may not be completely reliable or cannot be controlled at each stage for parameter inference, but with proper attributes and suitable architecture, an ML approach can be an alternative to existing methods in performing parameter estimation. In particular, the variational aspect of the conditional variational autoencoder \citep{2013arXiv1312.6114K,NIPS2015_8d55a249,Kingma_2019,doersch2021tutorialvariationalautoencoders,Gabbard_2021,Bond_Taylor_2022} (CVAE) transpires to be a suitable technique to carry out parameter inference. The structure of CVAE is composed of an encoder, a lower-dimensional latent space, and a decoder. During the training process in a CVAE, the encoder reduces the higher-dimensional input into lower-dimensional data and distributes it over the latent space. This distribution can be conditioned on the required features. The distributed data points in the latent space can be mapped back to the true data, and simultaneously, samples can be drawn for new parameters from this distribution. These drawn samples inherit the variational inference aspect, which is reflected in the variations among newly generated data points from one another even when sampled for the same parameter set. \par

The adaptability in the CVAE approach for parameter inference lies in the ability to accommodate data from various models in astrophysics and astronomy in the form of images, numerical arrays, hierarchical data structures, or other relevant data types for training and, consequently, generating the required physical parameters irrespective of the data type and resolution. Factors such as multimodality, hierarchical structures, non-gaussianity, degeneracies, etc. often lead to complex posterior distributions. Even for such specific and complex models, CVAE can be tuned seamlessly to capture the high-dimensional posterior distribution, even handling the uncertainty. \par

Kilonova \citep{Li_1998,  1999A&A...341..499R,2010MNRAS.406.2650M,Lipunov_2017}, powered by the radioactive decay in the merger ejecta and emitted in the form of electromagnetic waves, has been a topic of keen interest since its discovery in 2017 through the merger of the binary neutron star (BNS)\citep{2017ApJ...848L..16S,Lipunov_2017,Tanvir_2017,Arcavi_2017,2017ApJ...848L..24V,doi:10.1126/science.aap9811}, popularly known as GW170817. Even though there is another KNe combined with gravitational waves yet to be discovered, the emission mechanism and the final remnant of the BNS merger have intrigued research groups to delve deeper into it, consequently updating our knowledge about the physics behind KNe emission. In this work, we have demonstrated CVAE's ability in rapid and accurate KNe parameter inference by training the neural network with KNe light curve data while conditioning on the model specific physical parameters. Even though, in this paper, we have mainly focused on two particular data sets, other similar data sets with KNe features can be incorporated to achieve specific results. While generating required features, the CVAE is flexible and independent of the resolution of the light curve data, hence providing a nuanced approach for future observations. Therefore, in the occurrence of a future observation, even if the KNe light curve or spectra have a low resolution or fewer data points, the CVAE can be entrusted to generate a wide range of features within accepted error ranges. Multiple KNe models can be trained with the CVAE to develop and adopt a generalized approach while generating parameters specific to the trained models; therefore, for a particular light curve, the trained CVAE can specifically produce parameters that vary with the KNe models, which would indeed provide a wider understanding of the merging system. The CVAE architecture can be equally trained on light curves or spectra to obtain the distribution of physical parameters, which makes it versatile for multiple outputs.

\section{Data}
\label{sec:data}

In this section, we have highlighted the two different KNe data sets that have been used for training and inference. The respective KNe models, physical parameters, and their values have been explained. For the first data set, hereafter D$_{\mathrm{A}}$, the light curves, KNe models, and other specifications have been adopted from \citet{2019PhRvD.100b3008M,Korobkin_2021,2021ApJ...918...10W}. There are $900$ light curves that have a unique set of physical parameters. These are further classified according to the higher and lower electron fractions into $wind\_1(Y_e=0.37)$ and $wind\_2(Y_e=0.27)$, having $450$ light curves for each category. All these $450$ light curves have a unique combination of physical parameters with respect to the mass($M_{\odot}$) and velocity ($c$) of the high and low $Y_e$, respectively. For the masses of the high and low $Y_e$, we will follow the same nomenclature and term these \textit{md} and \textit{mw}, while for the velocities, it will be \textit{vd} and \textit{vw} , respectively. Throughout the text and subsequent plots, these abbreviations will be used wherever necessary. Furthermore, the high and low $Y_e$ are also morphologically categorized into Toroidal-Spherical (\textit{TS}) and Toroidal-Peanut (\textit{TP}) respectively. Hence, this categorization results in the TS and TP groups that have $wind\_1$ and $wind\_2$ models, and for each TS-$wind\_1$ and TP-$wind\_1$, there are $225$ light curves having $225$ unique combinations of the physical parameters based on the values of \textit{md,vd,mw} , and \textit{vw}. A similar classification is performed for TS-$wind\_2$ and TP-$wind\_2$ equally. 
.For the physical parameters, \textit{md} and \textit{mw} are characterized by values in the range ${[0.001, 0.003, 0.01, 0.03, 0.1]} M_{\odot}$, while \textit{vd} and \textit{vw} take values of ${[0.05, 0.15, 0.3]}c$, providing a complete set of configurations for the mass and velocity profiles. Throughout the paper, we strictly follow the same model name formats to describe the physical parameters as mentioned in \cite{2021ApJ...918...10W}. In addition, for each set of physical parameters, the light curves are segregated into $54$ angular bins that are uniform in solid angle. For each light curve across the $54$ angular bins, the set of physical parameters remains the same. Thus, for each KNe model, we have $12150$ light curves consistent with those physical parameters. From the above data, we randomly choose the physical parameters while splitting into $175$, $25$, and $25$ for training, testing, and validation, respectively. Moreover, these light curves are available in \textit{g,r,i,z} , and \textit{y} bands, enhancing the breadth of information available for multiwavelength analysis. Details of the models, physical parameters, and light curves are included in Table~\ref{table:1}.   \par

To generalize and provide a robust proof for the application of the proposed method, we execute the CVAE on another KNe data, hereafter D$_{\mathrm{B}}$, set adopted from \url{https://github.com/mnicholl/kn-models-nicholl2021}\citep{2021MNRAS.505.3016N}. For this data, we have the chirp mass($M_{\odot}$), mass ratio, fraction of the remnant disk, and the viewing angle (in degrees) as the physical parameters alongside the light curves in the filter bands of \textit{g,r,y, i, and z}. Table \ref{table:matt_table} highlights the physical parameters and their subsequent values. For each filter band, there are 513 light curves corresponding to different combinations of physical parameters—chirp mass, mass ratio, fraction of the remnant disk, and viewing angle—as listed in Table~\ref{table:matt_table}. The detailed physical parameters for each light curve can be obtained from the data link provided above. The use of two independent datasets allows us to assess the general applicability of the CVAE framework for inferring model-dependent physical parameters.

\begin{table}[ht!]
\centering
\begin{tabular}{lccc}
\hline\hline
Kilonova Model & Morphology & No. of Parameters\tablenotemark{a} & No. of Light Curves \\
\hline
\multirow{2}{*}{wind\_1 ($Y_e=0.37$)} 
  & Toroidal-Peanut    & 225 & 12150 \\
  & Toroidal-Spherical & 225 & 12150 \\
\hline
\multirow{2}{*}{wind\_2 ($Y_e=0.27$)} 
  & Toroidal-Peanut    & 225 & 12150 \\
  & Toroidal-Spherical & 225 & 12150 \\
\hline
\end{tabular}
\tablenotetext{a}{Physical parameters: $m_d$, $v_d$, $m_w$, $v_w$.}
\caption{For the data set D$_{\mathrm{A}}$, this table provides the models, number of physical parameters, and the number of light curves used during training, validation, and testing. Each model has $225$ physical parameters, where each parameter corresponds to a single light curve. For a single set of parameters, there are $54$ angular bins uniform in solid angle, amounting to $12150$ light curves for each KNe model. From these sets of parameters, we randomly split into training, test, and validation sets with $175$, $25$, and $25$ parameters, respectively. For md and mw, the physical parameters have values corresponding to ${[0.001,0.003,0.01,0.03,0.1]}M_{\odot}$ while for vd and vw the values are ${[0.05,0.15,0.3]}c$. The light curves are provided across multiple bands, specifically  in \textit{g,r,i,z} , and \textit{y} bands. }
\label{table:1}
\end{table}

\begin{table*}[ht!]
  \centering
  \renewcommand{\arraystretch}{2}
  \resizebox{7in}{!}{
    \begin{tabular}{|c|c|c|c|}
      \hline
      \textbf{Chirp Mass ($M_{\odot}$)} & 
      \textbf{Mass Ratio} & 
      \textbf{Remnant Disk Fraction} & 
      \textbf{Viewing Angle} \\
      \hline
       $1.0$, $1.2$, $1.4$, $1.6$, $1.8$ &
       $0.7$, $0.75$, $0.8$, $0.85$, $0.9$ &
       $0.15$, $0.20$, $0.25$, $0.30$, $0.35$, $0.40$ &
       $45\degree$, $60\degree$, $75\degree$, $90\degree$ \\
      \hline
    \end{tabular}
  }
  \caption{For the data set D$_{\mathrm{B}}$,this table contains physical parameters and there respective ranges that were used as conditioning parameters, while being trained with the light curves.}
  \label{table:matt_table}
\end{table*}

\section{Methodology}
\label{sec:method}
The CVAE architecture, which combines primary input data and additional conditional information to encode and consequently reconstruct data in a probabilistic manner, implemented in this work, has two input layers, \texttt{input\_1} and \texttt{input\_2}, with shapes \texttt{(None,4)} and \texttt{(None, 261)} , respectively. While the former, which consists of the data that would be reconstructed, represents the auxiliary information of physical parameter types of the KNe model, the latter consists of the conditional input, KNe light curves, to guide the encoding, decoding, and reconstruction. To ensure that the encoder considers both the data and the condition, these inputs are concatenated (\texttt{concatenate}) into a single vector of shape \texttt{(None, 265)}. A series of fully connected dense layers are progressively utilized to reduce the dimensionality, further extracting the features that are to be encoded into the lower-dimensional latent space. The encoder gradually compresses the input dimension from $265$ to $256$ and then finally to $32$, generating a mean vector and log variance, each having a shape of \texttt{(None,2)}, which parameterizes the latent space distribution.

The center and spread of the distribution are represented by the mean and log variance, respectively, thus defining a Gaussian distribution for the latent space. During backpropagation, to introduce stochastic sampling, the reparameterization trick is applied via a \texttt{lambda} layer. It samples a latent variable from the Gaussian distribution. The output of the \texttt{lambda} layer represents the compressed information of the physical parameters and the light curves of the chosen KNe model. This two-dimensional latent space has the essential variability of the data, along with uncertainty. The decoder accepts the output of the \texttt{ lambda} layer and concatenates it with the conditional input to form a shape input \texttt{ (None, 263)}.  The sequential dense layers in the decoder reconstruct the input while increasing the dimensionality. The desired output is achieved by the physical parameters for the conditioned light curve. Although the generated samples are diverse and have variations, they capture the intrinsic features of the expected output.

\begin{figure*}[ht!]
\centering
\includegraphics[width=\linewidth]{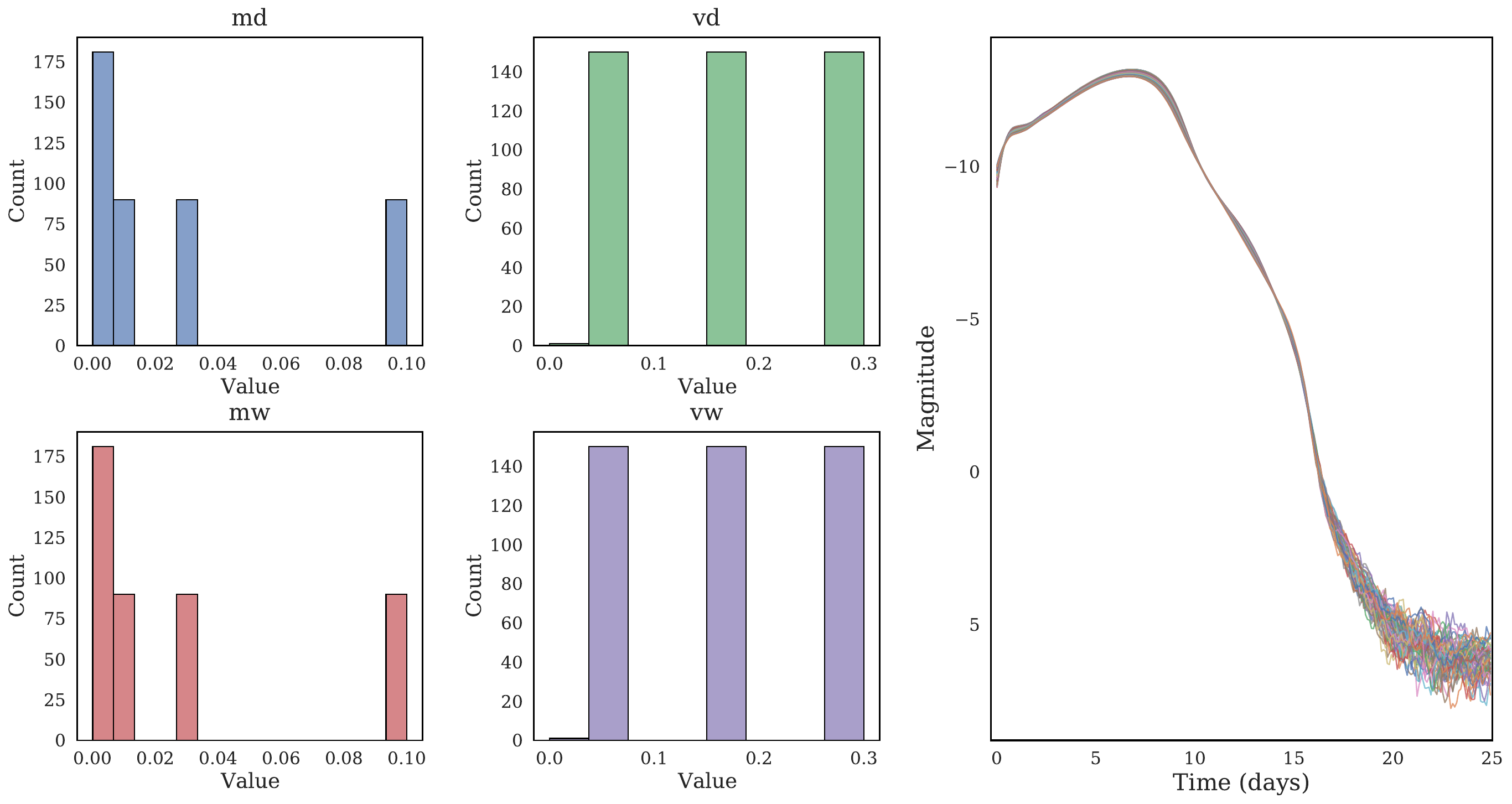}
\caption{In this plot, corresponding to the data set D$_{\mathrm{A}}$, the histogram depicts the distribution of the physical parameters that has been used for the training, testing and validation. As is evident, in the sets of physical parameters, we have only specific values. The light curve shown in the g band, is a sample light curve and has a single physical parameter set with \textit{md=$0.001M_{\odot}$}, \textit{vd=$0.05c$}, \textit{mw=$0.001M_{\odot}$} and \textit{vw=$0.05c$} in $54$ angular bins that are uniform in solid angle.}
\label{fig:com plot}
\end{figure*}

Ensuring that the main objective is parameter inference, we describe the generation of the parameter distribution for the light curves in the test data below. CVAE is trained separately on the different KNe models mentioned in Table~\ref{table:1}, while conditioning on the corresponding light curves. Upon successful training, while primarily checking the loss value during training and validation, we gain insight into the performance of the CVAE and consequently tune the hyperparameters to obtain better results. As mentioned above, even though there are $54$ different angular bins, during training, we include all these angular bins without explicitly stating the angular information in the CVAE. However, since the training is based on physical parameters conditioned on the light curves, while generating the physical parameters, we are agnostic about the angular-bin information and do not specify it in the results. Even though it is possible to strictly enumerate the angular bins while training and generating the physical parameters, to avoid cluttered plotting, we refrain from including those in the plots.  

The intrinsic variability introduced in the latent space provides the flexibility to generate any number of physical parameters for a single targeted light curve; however, in the current framework, we choose to generate a set of $1000$ physical parameters corresponding to \textit{md, vd, mw} , and \textit{vw} for individual light curves of the test set, which are further averaged while including the results in the plot. Similar methodology is applied on the hereafter D$_{\mathrm{B}}$ to recover the physical parameters. In addition, we have applied the trained CVAE on the observed KNe (AT 2017gfo) light curves to recover the chirp mass, mass ratio, fraction of the remnant disk, and viewing angle. After recovering these physical parameters, we have compared them with the values provided in \citet{2021MNRAS.505.3016N}. To determine the precision between the true and CVAE-generated physical parameters, we calculate the median, interquartile range (IQR), and mean squared error (MSE) between these parameters; however, as expected, there are slight variations among the median, IQR, and MSE values. We perform this analysis specifically for D$_{\mathrm{A}}$, as a direct one-to-one correspondence between the physical parameters of the model and those inferred from the observed KNe is not readily available. With the given data, training is completed within $\sim1$ hours, while generating $1000$ physical parameter sets across the models, covering all the filter bands, takes $\sim 10$ milliseconds. It is crucial to mention that training, validation, and testing are carried out using the \texttt{CPU} only. In the following section, we have included the important results obtained after the implementation of the CVAE on the mentioned data structure.

\section{Results}

As there is a paucity of a single metric that would completely bolster the CVAE-generated results, we will utilize a combination of violin plots, bandwise kernel density approximation (KDE) plots, polar plots in multiple bands, and mean squared error (MSE) represented via bar graphs for each type of physical parameter to verify the accuracy of the CVAE-generated physical parameters across the entire parameter space. For the main part of the article, the plots corresponding to \textit{TP-wind\_1} are shown, and for the remaining models, the plots are included in the appendix. In the following plots, we compare the distribution of the CVAE-generated parameters and true parameters across the different KNe models, adding the filter bands to depict more generalized results. For the violin and KDE plots, we will show the results from the g-band data only, while for the polar plots, all bands are utilized to show the results. In Fig.~\ref{fig:com plot}, we show the distribution of the physical parameters and a sample light curve that has \textit{md=$0.001M_{\odot}$}, \textit{vd=$0.05c$}, \textit{mw=$0.001M_{\odot}$} , and \textit{vw=$0.05c$}. As evident from the plot, the physical parameter distribution is not continuous and has discrete values; hence, it does not cover the entire parameter space. For the shown light curve, it ends approximately around $25$ days, where the variations indicate the simulated values associated with the $54$ angular-bin.  \par

\begin{figure*}[ht]
\centering
\plotone{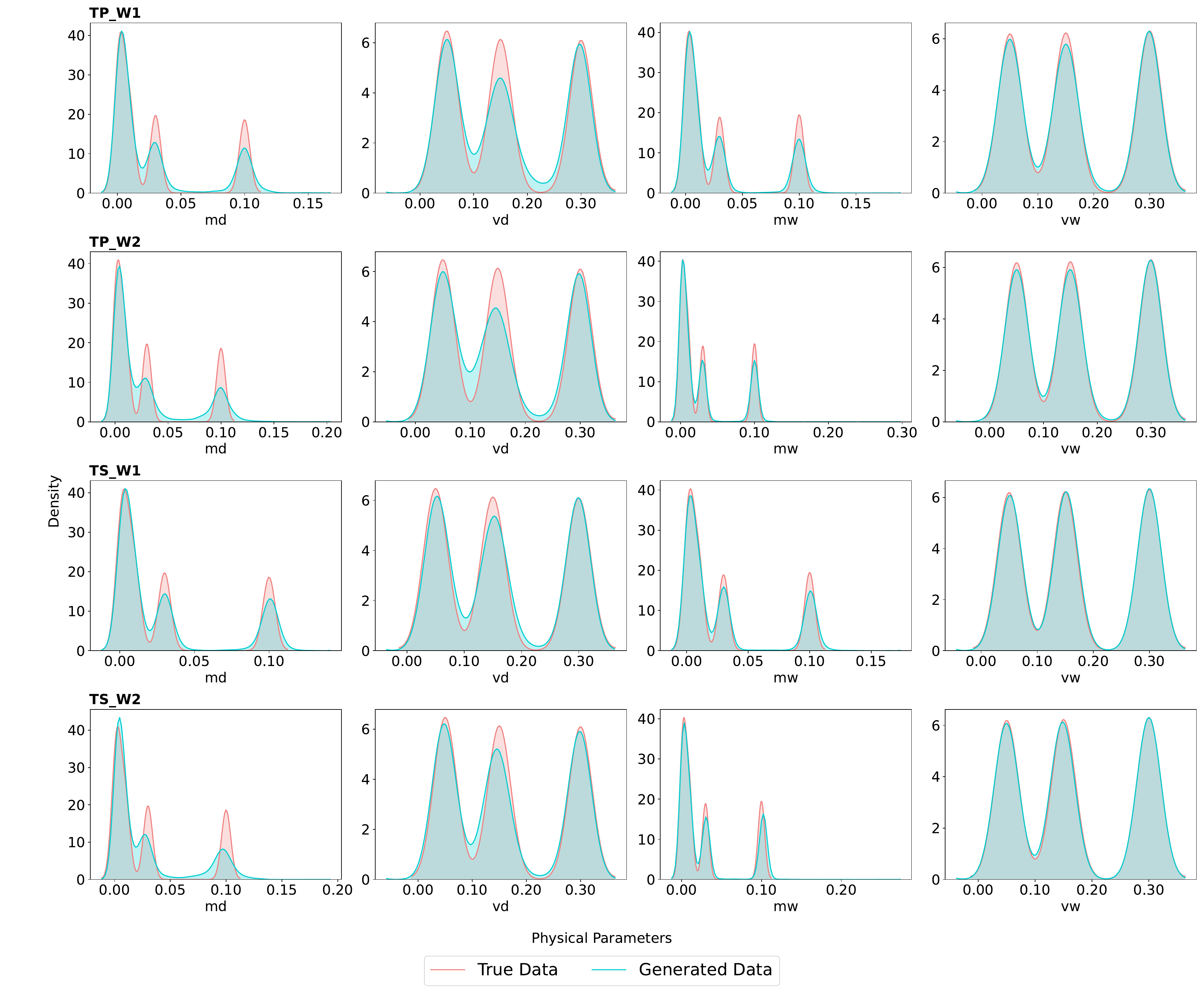}
\epsscale{1.3}
\caption{The plot outlines the kernel density estimation  between the true and CVAE-generated physical parameters for \textit{TP-wind\_1},\textit{TP-wind\_2},\textit{TS-wind\_1} and \textit{TS-wind\_2} models. Primarily, g-band light curves from the test data have been used to generate these physical parameters. The true and CVAE-generated distribution across the models for each of the physical parameters overlaps with each other hence representing a robust result obtained from the CVAE. However, in certain region of the parameter space, there are slight disagreement which can be attributed to the variational inference of the CVAE itself. }
\label{fig:KDE plot}
\end{figure*}

One of the potent applications for CVAE-generated parameter distributions would be to cover the entire physical parameter space that maps to light curves. The kernel density estimation plots shown in Fig.~\ref{fig:KDE plot} provide in-depth details about the overlap of CVAE-generated and true physical parameters for the test set g-band light curves across all models. It has been found that for certain values of physical parameters, the recovery is not appealing, and we suspect a lack of training data in those parameter regions, which results in poor retrieval. In addition, the variational inference from the CVAE can also lead to such disagreements; however, the overall distribution of the true and CVAE-generated physical parameters overlaps with each other, thus depicting an accurate performance of the CVAE.While calculating the KDEs, since [\texttt{kernel=gaussian}] has been used, as expected, the estimated KDEs have slightly extended into the negative region. However, since we are looking into the regions of superposition between the true and generated values, these extensions into the negative region are ignored.
In Table~\ref{tab:median_IQR} and Table~\ref{tab:IQR_only}, we have tabulated the median and IQR values, respectively, between the true and CVAE-generated physical parameters for all filter bands across the KNe models. For most of the physical parameters across the filter bands and models, we do find the true and generated values to be in close proximity to each other. From the IQR values across the filter bands and physical parameters, as these values are compared to each other, we are assured that the neural network has been able to properly capture the underlying distribution between the true and generated data. This provides a conclusive benchmark for assessing CVAE performance. However, with more training data, recovery can achieve better results in the underrepresented regions. 
\par

\begin{figure*}[ht!]
\centering
\includegraphics[width=0.98\linewidth]{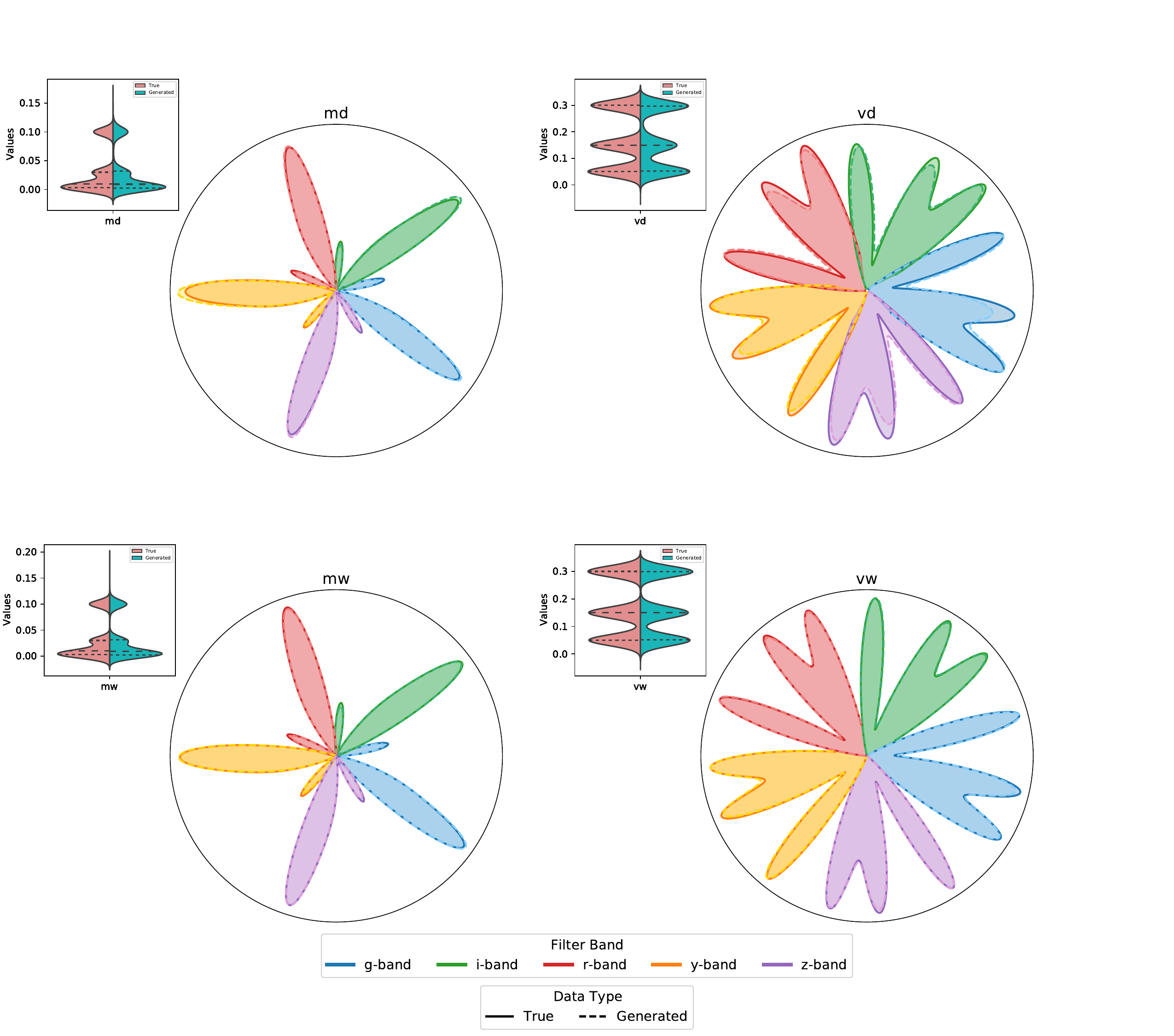}
\epsscale{1.0}
\caption{
This polar plot represents the kernel density approximation of the true and CVAE-generated physical parameters calculated in each filter band across the whole parameter space for md, vd, mw and vw respectively for \textit{TP-wind\_1} model. The different filter bands are shown along the angular axis. The violin plots in the inset corresponding to each polar plot compare the distribution of the true and CVAE-generated physical parameters specifically for the g-band.}
\label{fig:polar_plot_dense256}
\end{figure*}

In Fig.~\ref{fig:KDE plot}, only the g-band light curves of the test data have been used to generate the physical parameters; however, to incorporate a more general understanding, in Fig.~\ref{fig:polar_plot_dense256}, the band-wise distribution between the true parameters and those generated by CVAE has been shown for \textit{TP-wind\_1}.
Although the true physical parameters are defined only on a discrete parameter grid, we apply KDE in Fig.~\ref{fig:KDE plot} and Fig.~\ref{fig:polar_plot_dense256}  solely to obtain a smoothed representation of their empirical distribution for visual comparison with the CVAE-generated physical parameters. The apparent continuity of the generated distribution arises naturally from the continuous mapping learned by the CVAE between the latent, conditional, and output spaces.
In Fig.~\ref{fig:polar_plot_dense256} the KDE lobes are oriented along different filter bands. Each lobe, corresponding to the physical parameter, represents the comprehensive recovery of the physical parameter values over the entire parameter space in the test set. Since, for each type of physical parameter, we observe some discrete distributions along specific parameter values, similar features are also evident from the multiple lobes of the KDEs. From this figure, by looking at the overlap of the true lobes and the CVAE-generated lobes, we can confirm that the recovery of the physical parameters is quite satisfactory. Similar polar plots for other models are provided in the supplementary material. 

\par
\begin{figure*}[ht!]
\centering
\includegraphics[width=0.98\linewidth]{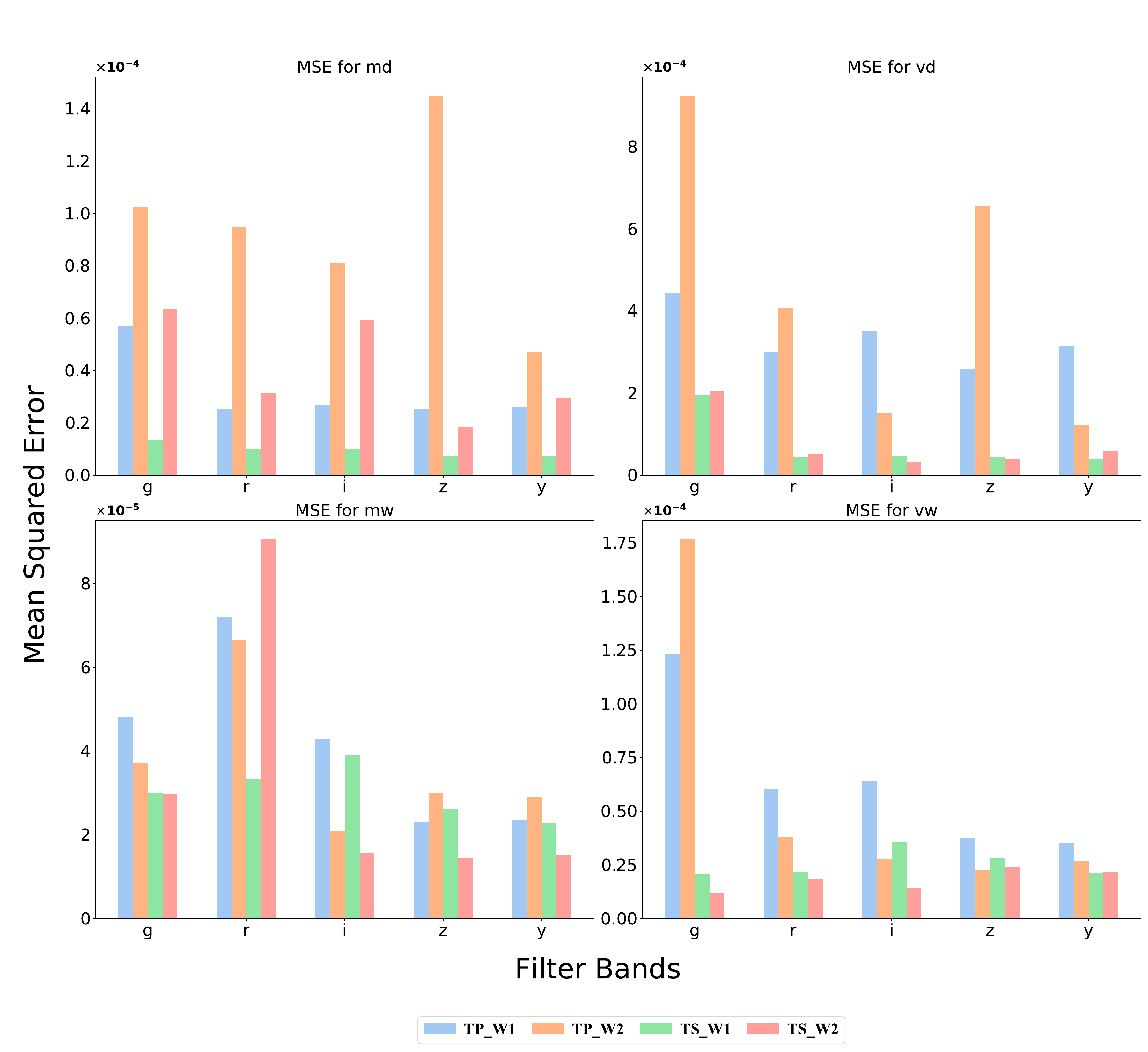}
\epsscale{1.0}
\caption{ Mean squared error plot for all physical parameters grouped according to the models for all the filter bands. In the consecutive frames shows the MSE for \textit{md}, \textit{vd}, \textit{mw} and \textit{vw} where in each frame, MSEs are grouped according to the KNe models across the filter bands.  
}
\label{fig:mse_all_models_all_filter}
\end{figure*}

To further investigate CVAE performance, in Fig.~\ref{fig:mse_all_models_all_filter} we plot the mean squared error between the true and CVAE generated values for all filter bands grouped according to the model across physical parameters. The MSE values are plotted along the y-axis for each frame, while the corresponding filter bands are shown along the x-axis. The MSE values of \textit{md},\textit{vd},\textit{mw}, and \textit{vw} are illustrated in \textit{the top left}, \textit{top right}, \textit{bottom left}, and \textit{bottom right} frames, respectively. For each frame, the MSEs are represented in the form of a bar graph grouped according to the models mentioned in Table~\ref{table:1}. As is evident, we have obtained a low value of MSE for all physical parameters, which resembles the accuracy among the generated values. During calculation, the average of the MSE across the physical parameters for the filter results is $\sim 0.00007$, but for \textit{vd}, we see large bars compared to other physical parameters; however, the MSE is accurate $\sim0.0001$. Now, comparing this deviation of MSE for \textit{vd} with the second plot from the left of the top panel in Fig.~\ref{fig:KDE plot} and the top right frame of Fig.~\ref{fig:polar_plot_dense256}, we particularly see some consistencies in the poor recovery of \textit{vd} , which could be attributed to a lack of training data in that region of parameter space. However, we expect that, with the addition of more training data spanning a wider and denser range of \textit{vd} values, the CVAE can achieve better inference accuracy.

\begin{table}[ht]
\large
\centering
\begin{tabular}{@{}l c cc cc cc cc@{}}
\toprule
Filter Bands & Model & \multicolumn{2}{c}{md} & \multicolumn{2}{c}{vd} & \multicolumn{2}{c}{mw} & \multicolumn{2}{c}{vw} \\
\cmidrule(lr){3-4} \cmidrule(lr){5-6} \cmidrule(lr){7-8} \cmidrule(lr){9-10}
 & & True & Generated & True & Generated & True & Generated & True & Generated \\
\midrule

\multirow{4}{*}{g-band}  
& TP\_W1 & 0.0100 & 0.0142 & 0.1500 & 0.1537 & 0.0100 & 0.0133 & 0.1500 & 0.1507 \\
& TP\_W2 & 0.0100 & 0.0096 & 0.1500 & 0.1301 & 0.0100 & 0.0099 & 0.1500 & 0.1543 \\
& TS\_W1 & 0.0100 & 0.0101 & 0.1500 & 0.1520 & 0.0100 & 0.0102 & 0.1500 & 0.1516 \\
& TS\_W2 & 0.0100 & 0.0094 & 0.1500 & 0.1450 & 0.0100 & 0.0101 & 0.1500 & 0.1485 \\
\midrule
\multirow{4}{*}{i-band}  
& TP\_W1 & 0.0100 & 0.0113 & 0.1500 & 0.1459 & 0.0100 & 0.0103 & 0.1500 & 0.1514 \\
& TP\_W2 & 0.0100 & 0.0103 & 0.1500 & 0.1476 & 0.0100 & 0.0101 & 0.1500 & 0.1516 \\
& TS\_W1 & 0.0100 & 0.0097 & 0.1500 & 0.1454 & 0.0100 & 0.0104 & 0.1500 & 0.1514 \\
& TS\_W2 & 0.0100 & 0.0096 & 0.1500 & 0.1498 & 0.0100 & 0.0097 & 0.1500 & 0.1499 \\
\midrule
\multirow{4}{*}{r-band}  
& TP\_W1 & 0.0100 & 0.0104 & 0.1500 & 0.1509 & 0.0100 & 0.0104 & 0.1500 & 0.1503 \\
& TP\_W2 & 0.0100 & 0.0104 & 0.1500 & 0.1506 & 0.0100 & 0.0102 & 0.1500 & 0.1512 \\
& TS\_W1 & 0.0100 & 0.0098 & 0.1500 & 0.1508 & 0.0100 & 0.0104 & 0.1500 & 0.1502 \\
& TS\_W2 & 0.0100 & 0.0100 & 0.1500 & 0.1522 & 0.0100 & 0.0100 & 0.1500 & 0.1501 \\
\midrule
\multirow{4}{*}{y-band}  
& TP\_W1 & 0.0100 & 0.0095 & 0.1500 & 0.1502 & 0.0100 & 0.0096 & 0.1500 & 0.1513 \\
& TP\_W2 & 0.0100 & 0.0095 & 0.1500 & 0.1498 & 0.0100 & 0.0092 & 0.1500 & 0.1503 \\
& TS\_W1 & 0.0100 & 0.0091 & 0.1500 & 0.1470 & 0.0100 & 0.0096 & 0.1500 & 0.1501 \\
& TS\_W2 & 0.0100 & 0.0093 & 0.1500 & 0.1500 & 0.0100 & 0.0087 & 0.1500 & 0.1508 \\
\midrule
\multirow{4}{*}{z-band}  
& TP\_W1 & 0.0100 & 0.0097 & 0.1500 & 0.1509 & 0.0100 & 0.0095 & 0.1500 & 0.1506 \\
& TP\_W2 & 0.0100 & 0.0099 & 0.1500 & 0.1538 & 0.0100 & 0.0096 & 0.1500 & 0.1503 \\
& TS\_W1 & 0.0100 & 0.0096 & 0.1500 & 0.1550 & 0.0100 & 0.0099 & 0.1500 & 0.1498 \\
& TS\_W2 & 0.0100 & 0.0095 & 0.1500 & 0.1498 & 0.0100 & 0.0097 & 0.1500 & 0.1500 \\

\bottomrule
\end{tabular}
\caption[Median values for true and CVAE-generated physical parameters]{This table provides the calculated values of median for the true and generated physical parameters across all the filter bands, categorized by KNe models.}
\label{tab:median_IQR}
\end{table}
\begin{table}[ht]
\large
\centering
\begin{tabular}{@{}l c cc cc cc cc@{}}
\toprule
Filter Bands & Model & \multicolumn{2}{c}{md} & \multicolumn{2}{c}{vd} & \multicolumn{2}{c}{mw} & \multicolumn{2}{c}{vw} \\
\cmidrule(lr){3-4} \cmidrule(lr){5-6} \cmidrule(lr){7-8} \cmidrule(lr){9-10}
& & True & Generated & True & Generated & True & Generated & True & Generated \\
\midrule

\multirow{4}{*}{g-band}
& TP\_W1 & 0.0270 & 0.0291 & 0.2500 & 0.2445 & 0.0270 & 0.0290 & 0.2500 & 0.2471 \\
& TP\_W2 & 0.0270 & 0.0299 & 0.2500 & 0.2448 & 0.0270 & 0.0290 & 0.2500 & 0.2480 \\
& TS\_W1 & 0.0270 & 0.0287 & 0.2500 & 0.2454 & 0.0270 & 0.0290 & 0.2500 & 0.2475 \\
& TS\_W2 & 0.0270 & 0.0273 & 0.2500 & 0.2482 & 0.0270 & 0.0294 & 0.2500 & 0.2495 \\
\midrule

\multirow{4}{*}{i-band}
& TP\_W1 & 0.0270 & 0.0286 & 0.2500 & 0.2424 & 0.0270 & 0.0297 & 0.2500 & 0.2480 \\
& TP\_W2 & 0.0270 & 0.0292 & 0.2500 & 0.2459 & 0.0270 & 0.0288 & 0.2500 & 0.2486 \\
& TS\_W1 & 0.0270 & 0.0276 & 0.2500 & 0.2491 & 0.0270 & 0.0302 & 0.2500 & 0.2477 \\
& TS\_W2 & 0.0270 & 0.0268 & 0.2500 & 0.2480 & 0.0270 & 0.0287 & 0.2500 & 0.2480 \\
\midrule

\multirow{4}{*}{r-band}
& TP\_W1 & 0.0270 & 0.0289 & 0.2500 & 0.2417 & 0.0270 & 0.0300 & 0.2500 & 0.2480 \\
& TP\_W2 & 0.0270 & 0.0282 & 0.2500 & 0.2401 & 0.0270 & 0.0289 & 0.2500 & 0.2487 \\
& TS\_W1 & 0.0270 & 0.0279 & 0.2500 & 0.2468 & 0.0270 & 0.0299 & 0.2500 & 0.2483 \\
& TS\_W2 & 0.0270 & 0.0284 & 0.2500 & 0.2456 & 0.0270 & 0.0284 & 0.2500 & 0.2486 \\
\midrule

\multirow{4}{*}{y-band}
& TP\_W1 & 0.0270 & 0.0277 & 0.2500 & 0.2394 & 0.0270 & 0.0303 & 0.2500 & 0.2467 \\
& TP\_W2 & 0.0270 & 0.0289 & 0.2500 & 0.2414 & 0.0270 & 0.0300 & 0.2500 & 0.2471 \\
& TS\_W1 & 0.0270 & 0.0279 & 0.2500 & 0.2474 & 0.0270 & 0.0290 & 0.2500 & 0.2481 \\
& TS\_W2 & 0.0270 & 0.0307 & 0.2500 & 0.2416 & 0.0270 & 0.0289 & 0.2500 & 0.2418 \\
\midrule

\multirow{4}{*}{z-band}
& TP\_W1 & 0.0270 & 0.0292 & 0.2500 & 0.2400 & 0.0270 & 0.0299 & 0.2500 & 0.2478 \\
& TP\_W2 & 0.0270 & 0.0276 & 0.2500 & 0.2427 & 0.0270 & 0.0295 & 0.2500 & 0.2486 \\
& TS\_W1 & 0.0270 & 0.0286 & 0.2500 & 0.2459 & 0.0270 & 0.0297 & 0.2500 & 0.2485 \\
& TS\_W2 & 0.0270 & 0.0275 & 0.2500 & 0.2456 & 0.0270 & 0.0291 & 0.2500 & 0.2474 \\

\bottomrule
\end{tabular}
\caption{Interquartile range calculated for the true and generated values across the KNe models for all filter bands.}
\label{tab:IQR_only}
\end{table}



\begin{figure*}[ht]
\centering
\includegraphics[width=\linewidth]{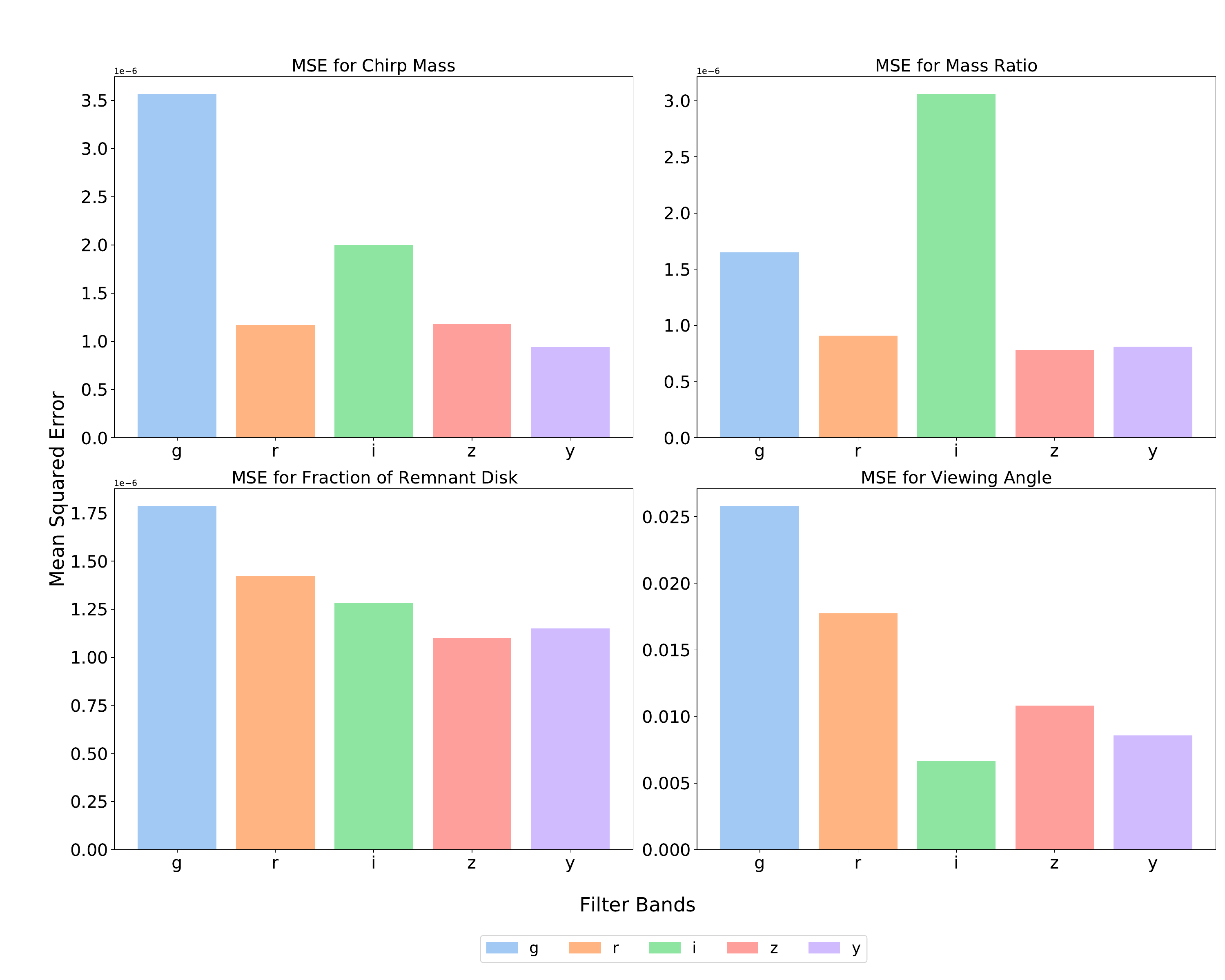}
\caption{This plot represents the mean squared error calculated between the true and CVAE-generated physical parameters for chirp mass, mass ratio, fraction of the remnant disk and viewing angle while grouped according to the filter bands.}
\label{fig:mse_matt_all_filter}
\end{figure*}

\begin{figure*}[ht]
\centering
\includegraphics[width=\linewidth]{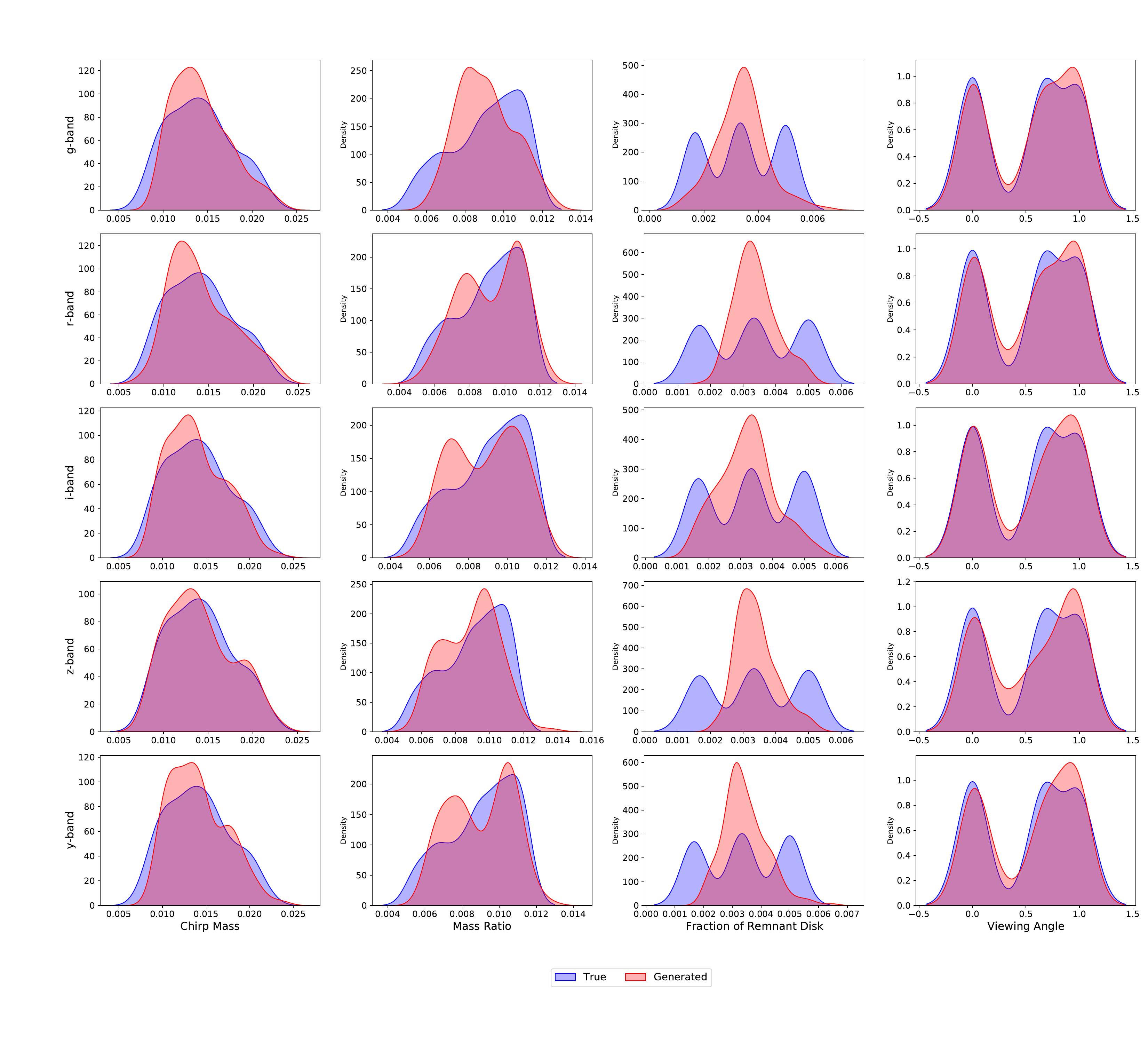}
\caption{Kernel density estimation plot for all physical parameters grouped according to the filter bands. This comprehensive plot demonstrates the extent of intersection between the true and generated physical parameters. In this subplot, down the rows, for each filter bands and across the rows for physical parameters, the KDEs for chirp mass, mass ratio, fraction of the remnant disk and viewing angle are depicted.}
\label{fig:kde_matt_all_filter}
\end{figure*}

 Keeping the training, testing, and validation split in the same ratio as in the earlier analysis, we perform parameter inference in a manner similar to that described in Section~\ref{sec:method}, by training the physical parameters while conditioning on the light curves. To evaluate the performance and maintain consistency, thus aiding in a comprehensive understanding, we have relied on kernel density estimation and calculated the mean squared error between the true and generated physical parameter sets.  Fig.~\ref{fig:mse_matt_all_filter} illustrates the calculated mean squared error between the true and CVAE-generated physical parameters across the filter bands. Obtaining the results for this data, the average mean squared error is of the order of \(10^{-4}\), calculated across all parameter space and filter bands; however, when calculated individually, it differs for the viewing angle, with a mean value of $0.015$ across all filter bands, which is expected as there are only three discrete values for the viewing angle. Additionally, the kernel density estimation across the filter bands for all the physical parameters is shown in Fig.~\ref{fig:kde_matt_all_filter}. From this figure, the consistencies between the true and CVAE-generated physical parameters are evident from the extent of superposition between the KDE lobes. As evident, the KDEs have acceptable superposition for chirp mass, mass ratio, and viewing angle; however, for the fraction of the remnant disk, the generated physical parameters are grouped along some discrete values. \par

For D$_{\mathrm{B}}$, the CVAE model is trained to recover the chirp mass, mass ratio, fraction of the remnant disk, and viewing angle. Thus, aiming to recover the corresponding physical parameters of GW170817, we apply the trained CVAE-model to the observed KNe data (\url{https://kilonova.org/data}). After inferring the physical parameters from the CVAE model, we compared those with the values detailed in \citet{2021MNRAS.505.3016N}. After applying the trained model, we obtained $1.204(1.188) M_{\odot}$, $0.974 (0.92)$, $0.151 (0.12)$ and $12\degree(\approx33\degree$), for chirp mass, mass ratio, fraction of the remnant disk, and viewing angle, where the values in the brackets are adopted from the agnostic values provided in Table 1 in \citet{2021MNRAS.505.3016N}. With the exception of the viewing angle, the CVAE accurately recovers the physical parameters. The limited accuracy in the inferred viewing angle is likely due to the discrete nature of the corresponding values in the training data. Utilizing a denser angular grid is expected to improve interpolation between adjacent viewing angles and should consequently reduce such discrepancies by allowing the CVAE to learn from a smooth posterior. Having more values of the viewing angles in the training set, would expose the CVAE to more gradual variations. This is preferable rather than being limited by a small number of discrete angular values. Consequently, this should mitigate the bias of the inferred viewing angle toward the specific angular values present in the training set, enabling the CVAE to recover with enhanced accuracy.
 Besides this limitation, the inferred parameter values demonstrate the robustness of the CVAE framework. It is important to note that, for the earlier case in D$_{\mathrm{A}}$, the trained CVAE-models lacked these physical parameters, thus making it comparatively unsuitable for subsequent inference.

\section{Discussion \& Conclusion}
In this paper, we investigate and present an alternative method for performing parameter inference using the Bayesian aspect of CVAE. The data used for training the CVAE has important features based on morphologies, making it suitable for covering a wider parameter space while simultaneously preserving astrophysical significance. The brevity shown by the CVAE architecture, after rigorous testing, has emerged as an alternative for generating desirable physical parameters. From the plots shown in the text, we can ascertain that the CVAE has consistently demonstrated promising results, illustrating that the potential of this technique to infer is not only suitable for KNe but can also be equally applicable to other transients. Chronologically, the plots represent the original distribution of the physical parameters alongside the outlook of the light curves, followed by KDE plots and violin plots for all the models, which provide conclusive evidence for obtaining accurate physical parameters, as is evident from the distribution across the filter bands. Furthermore, the MSE plots provide a comprehensive examination of the quality of the generated data not only across all the KNe models but also for the respective filter bands. We acknowledge that the current analysis is limited by the publicly available simulation set, which is defined only on a discrete parameter grid rather than over a densely sampled continuous space. For the discrete parameter grid in simulation data for training and inference, we expect to have the possibility of reduced accuracy or the introduction of minor systematic biases in areas of the parameter space that are sparsely sampled or close to the grid boundaries. Thus depending on the scale of reduction in accuracy, the impact on the inferred parameters can range from negligible to considerable. However, a comprehensive quantitative evaluation utilizing off-grid or continuously sampled parameter space is considered for future work.

Consequently, the present results should be interpreted as demonstrating the ability of the CVAE to learn the distribution of model light curves within the support of the available simulations, while the extension to a more continuous parameter space remains an important direction for future work. Indeed, upon introducing a larger data set, it is expected to enhance the CVAE’s ability to learn the mapping between light-curve features and posterior parameter distributions. This improvement would primarily stem from better coverage of sparsely sampled regions and more effective interpolation between neighboring grid points. Furthermore, larger training sets could reduce the model's sensitivity to individual examples, leading to smoother, more stable posteriors. However, the resulting performance gains will not simply scale with the increased training set. It would depend significantly on the training strategy employed, specifically if more data are focused on areas with rapid observable changes or pronounced parameter degeneracies. A systematic evaluation of both the training-set size and the optimal strategy is open for future work. Finally, we demonstrated that the trained CVAE can be applied to the observational data to recover physical parameters with high fidelity.
Most importantly, by applying this technique in a real-time scenario, parameter inferences can be carried out rapidly by providing only a light curve as input, thus generating the physical parameters.


\clearpage
\begin{acknowledgments}
This work is supported by the National Science and Technology Council of Taiwan under the grants 114-2112-M-007-033-MY3 and 114-2811-M-007-077. This work used high-performance computing facilities operated by the Center for Informatics and Computation in Astronomy (CICA) at National Tsing Hua University. This equipment was funded by the Ministry of Education of Taiwan, the Ministry of Science and Technology of Taiwan, and National Tsing Hua University.
\end{acknowledgments}

\software{\texttt{Matplotlib\citep{Hunter:2007}}, \texttt{scikit-learn}\citep{scikit-learn},\texttt{keras}\citep{chollet2015keras}, \texttt{tensorflow\citep{tensorflow2015-whitepaper}}, \texttt{seaborn \cite{Waskom2021}}}

\appendix

\section{Appendix}
\subsection{Additional Results}
\begin{figure*}[ht]
\centering
\includegraphics[width=\linewidth]{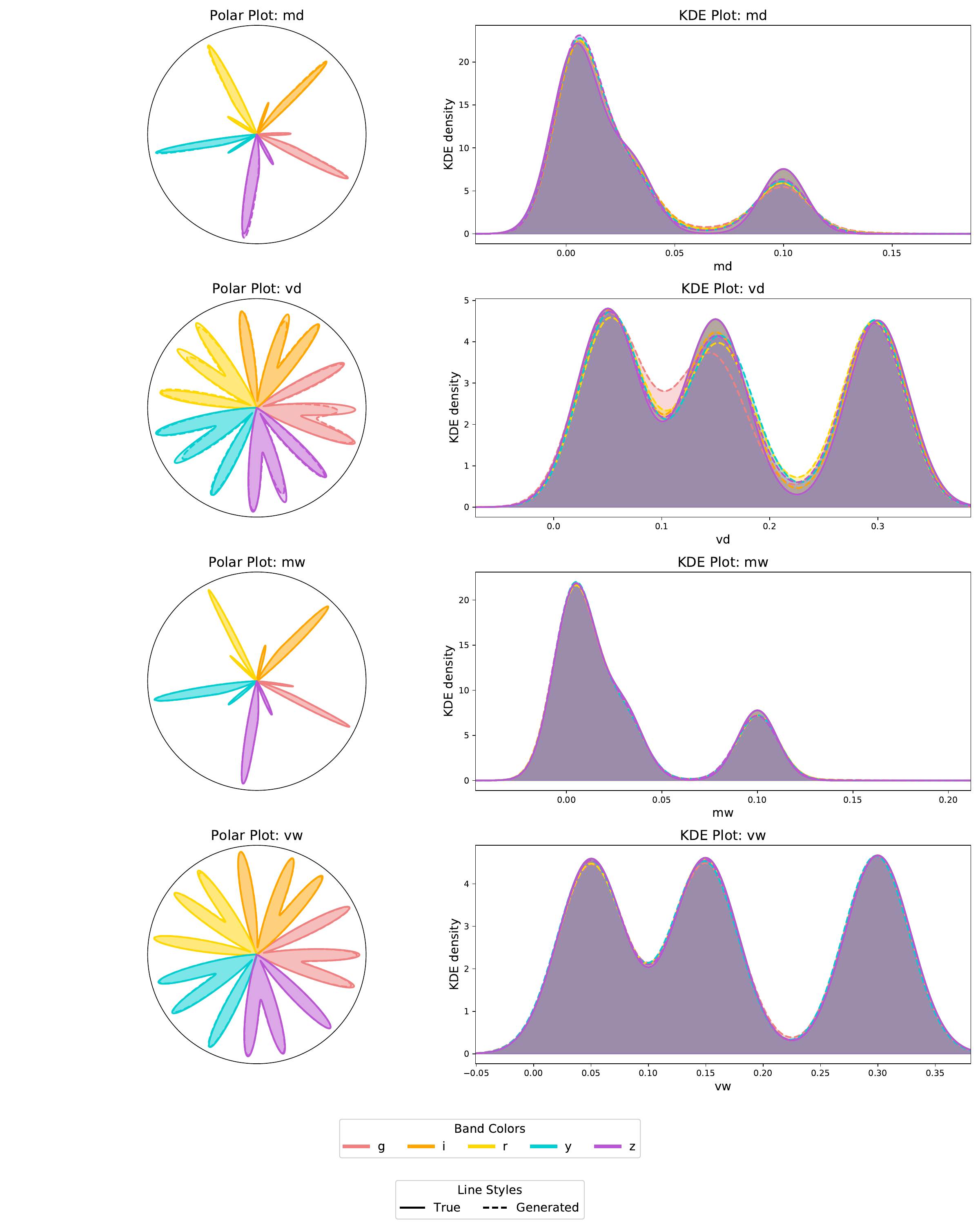}
\caption{Polar plot combined with KDE plot for all physical parameters across all filter bands for \textit{TP-wind\_2} model.}
\label{fig:polar_kde_tp_w2}
\end{figure*}

\begin{figure*}[ht]
\centering
\includegraphics[width=\linewidth]{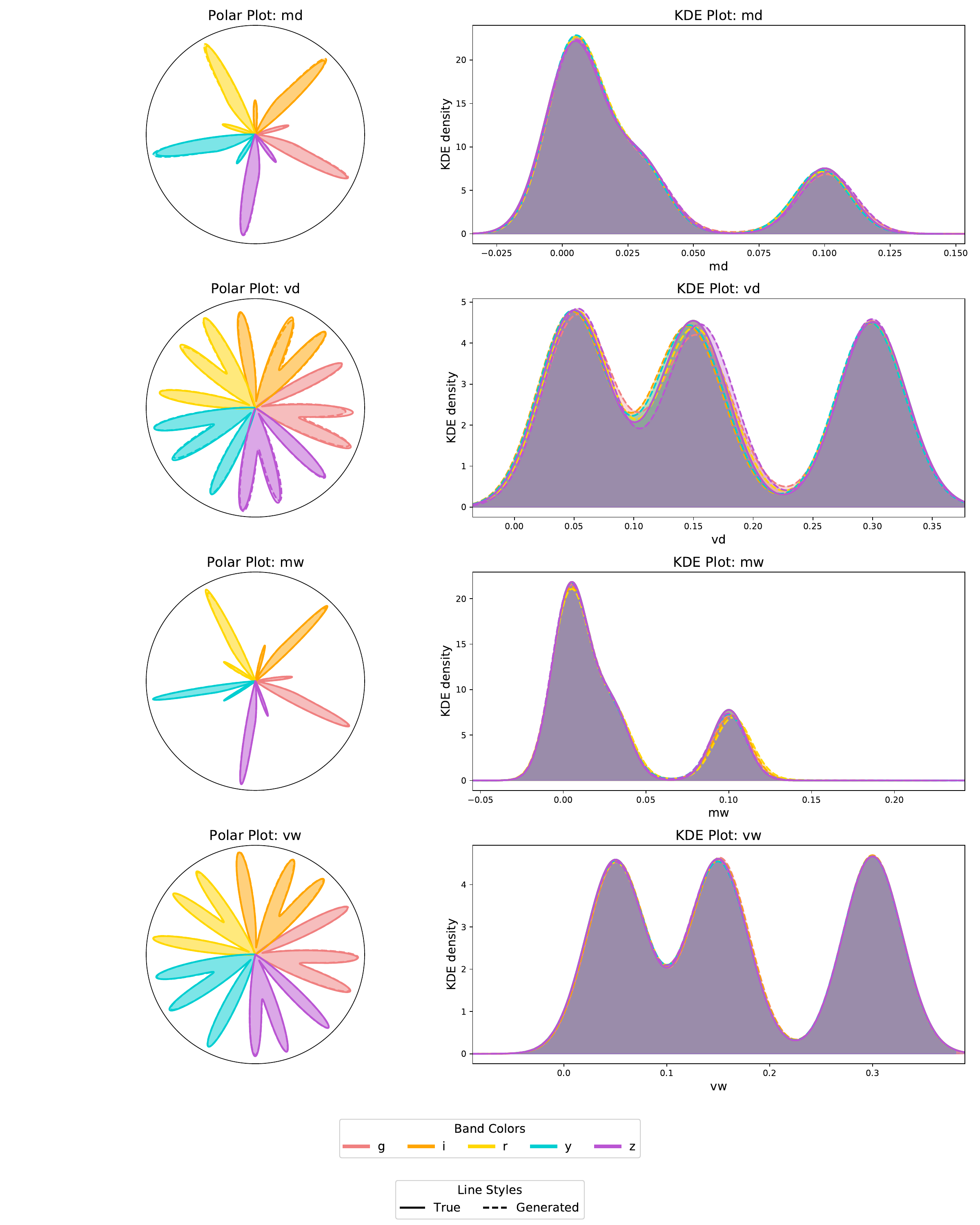}
\caption{Polar plot combined with KDE plot for all physical parameters across all filter bands for \textit{TS-wind\_1} model.}
\label{fig:polar_kde_ts_w1}
\end{figure*}

\begin{figure*}[ht]
\centering
\includegraphics[width=\linewidth]{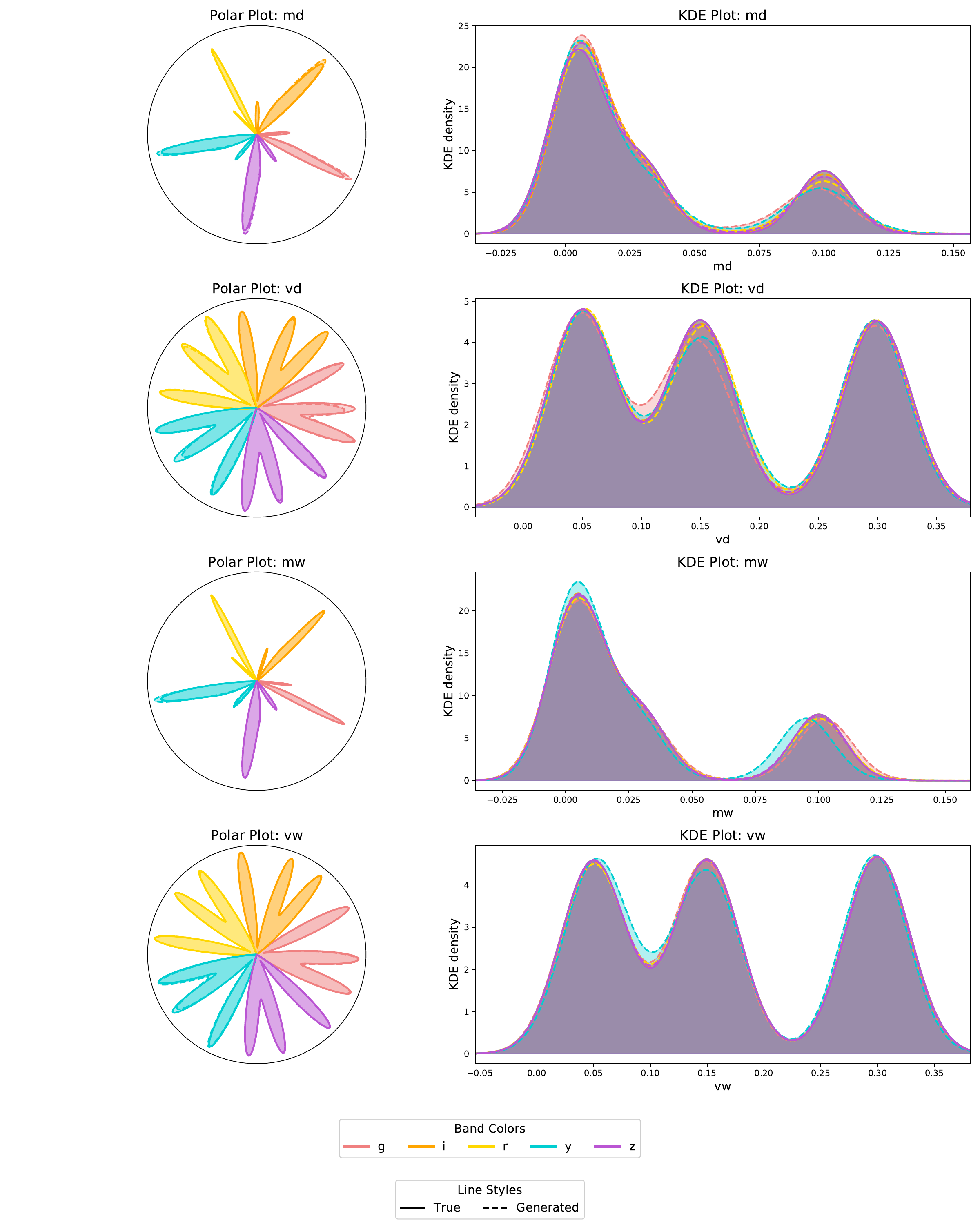}
\caption{Polar plot combined with KDE plot for all physical parameters across all filter bands for \textit{TS-wind\_2} model.}
\label{fig:polar_kde_ts_w2}
\end{figure*}

From Fig.~\ref{fig:polar_kde_tp_w2} to Fig.~\ref{fig:polar_kde_ts_w2}, we have shown the additional results after implementing the CVAE on the remaining KNe models. In these figures, the polar plot and the KDE plot are demonstrated across the different physical parameters in the filter bands. We do find similarly accurate results and hence conclude the notably robust performance of the CVAE. 

\subsection{Data Availability} For the former primary results of parameter inference, as mentioned in the main text, we have used the public data available at \url{https://zenodo.org/records/5711234} while for the latter additional results, we have relied on \url{https://github.com/mnicholl/kn-models-nicholl2021}. 

\clearpage
\nocite{*}
\bibliography{references}{}
\bibliographystyle{aasjournalv7}

\end{document}